\documentclass[aps,prl,reprint,superscriptaddress,amsmath,footinbib]{revtex4-1}
\usepackage{graphicx}
\usepackage{bbold}
\usepackage{hyperref}
\usepackage{import}
\usepackage{color}
\usepackage{soul}

\newcommand{\bra}[1]{\ensuremath{\langle #1|\,}}

\newcommand{\ket}[1]{\ensuremath{\,|#1\rangle}}
\newcommand{\Ca}{\ifmmode ^{40}\text{Ca}^{+} \else $^{40}$Ca$^{+}$~\fi}
\newcommand{\mus}{\ifmmode \mu\mathrm{s} \xspace \else $\mu$s \xspace\fi}

\newcommand{\executeiffilenewer}[3]{\ifnum\pdfstrcmp{\pdffilemoddate{#1}}%
{\pdffilemoddate{#2}}>0{\immediate\write18{#3}}\fi}

\begin{document}


\title{Heralded entanglement of two ions in an optical cavity} 



\author{B. Casabone}
\author{A. Stute}
\author{K. Friebe}
\author{B. Brandst\"atter}
\author{K. Sch\"uppert}
\affiliation{Institut f{\"u}r Experimentalphysik, Universit{\"a}t Innsbruck, Technikerstra{\ss}e 25, 6020 Innsbruck, Austria}
\author{R. Blatt}
\affiliation{Institut f{\"u}r Experimentalphysik, Universit{\"a}t Innsbruck, Technikerstra{\ss}e 25, 6020 Innsbruck, Austria}
\affiliation{Institut f\"ur Quantenoptik und Quanteninformation der \"Osterreichischen Akademie der Wissenschaften,
Technikerstra{\ss}e 21a, 6020 Innsbruck, Austria}
\author{T. E. Northup}
\email{tracy.northup@uibk.ac.at}
\affiliation{Institut f{\"u}r Experimentalphysik, Universit{\"a}t Innsbruck, Technikerstra{\ss}e 25, 6020 Innsbruck, Austria}
\date{\today}

\begin{abstract}
We demonstrate precise control of the coupling of each of two trapped ions to the mode of an optical resonator.
When both ions are coupled with near-maximum strength, we generate ion--ion entanglement heralded by the detection of two orthogonally polarized cavity photons.
The entanglement fidelity with respect to the Bell state $\Psi^+$ reaches $F \geq (91.9\pm2.5)\%$.
This result represents an important step toward distributed quantum computing with cavities linking remote atom-based registers.
\end{abstract}


\maketitle 

%
%

Key experiments have explored the interaction of single trapped atoms with an optical cavity mode \cite{Miller05,Ritter12},
a paradigmatic system that lends itself to the study of quantum processes.
One can also approach the atom--cavity system from the regime of large atom numbers and couple a single cavity mode to an ensemble of $N \gg 1$ atoms
\cite{Black03,Brennecke07,Colombe07,Herskind09}.
In this case, the coupling strength scales as $\sqrt{N}$,  and novel collective effects such as spatial self-organization and phase transitions can be observed \cite{Black03,Baumann10}.
However, in these experiments, information about the quantum states of individual atoms is not directly accessible.
By working with just a few trapped particles, one can take advantage of the degree of control available in single-atom experiments while exploring the richer physics of multi-atom interactions.
From the perspective of quantum information science, multiple atoms within a cavity can provide error correction in quantum networks \cite{vanEnk97}, improve quantum memories \cite{Lamata11}, and generate multi-dimensional cluster states \cite{Economou10}.
The precise positioning of an array of atoms with respect to the cavity mode is a prerequisite for
gates based on time-dependent interactions \cite{Beige00} and quantum simulations of the Bose-Hubbard and Frenkel-Kontorova models \cite{Johanning09}.

In this Letter, we report on coupling two ions to the mode of an optical cavity and show that the interaction strength of each ion with the cavity can be controlled.
Next, we demonstrate a protocol that relies on this coupling:  heralded entanglement between the two ions.
This cavity-based method represents a promising route for generating entanglement in quantum registers,
such as arrays of neutral atoms.
For ions confined in a shared potential as in our experiment, local entanglement is already possible via motional degrees of freedom \cite{Blatt08}.
However, our result constitutes a stepping stone for efficient entanglement of remote ions \cite{Moehring07,Maunz09}, distributed quantum computing \cite{Jiang07}, and protocols requiring the controlled coupling of multiple ions to a single cavity \cite{Lamata11,Economou10,Beige00,Johanning09}.

Entanglement has previously been demonstrated between two Rydberg atoms traversing a microwave cavity \cite{Hagley97,Osnaghi01}, based on the unitary evolution of the atom--cavity interaction at rate $2g$.
For high-fidelity entanglement, such schemes \cite{Pellizzari95,Zheng00} typically require the strong-coupling regime $g \gg \{ \kappa, \gamma \}$, where $\kappa$ and $\gamma$ are decay rates of the cavity field and the atom.
More recent strategies for dissipative preparation of entanglement are less stringent but still assume a cooperativity parameter $\mathcal{C} \equiv g^2/(2\kappa\gamma)$ of more than 10 \cite{Haerkoenen09,Kastoryano11,Busch11}.
Here, we use a modified version of the proposal by Duan and Kimble \cite{Duan03}, in which the entanglement fidelity is robust to spontaneous emission and the probability of success approaches 1/2.

Our experimental system consists of a linear Paul trap within an optical cavity in an intermediate coupling regime. 
The system parameters $\{g_{\text{eff}}, \kappa, \gamma_{\text{eff}} \}$ are $2 \pi \times \{37, 50, 54\}$~kHz, where the effective rates $g_{\text{eff}}$ and $\gamma_{\text{eff}}$ are determined by mapping a  three-level atomic system with a detuned drive field onto an effective two-level system \cite{Stute12a,SuppMat}.
The cavity axis is nearly orthogonal to the trap axis, along which strings of ions are confined.
In order to demonstrate the control that this system affords in coupling multiple ions to the cavity mode, we show two ion--cavity configurations.
In the first, one ion is maximally coupled to the cavity, and another ion is minimally coupled.
The second configuration corresponds to both ions maximally coupled to the cavity mode.

\begin{figure}
\def\svgwidth{\columnwidth}
\executeiffilenewer{figure1.svg}{figure1.pdf}{inkscape -z -D --file=figure1.svg --export-pdf=figure1.pdf --export-latex --export-area-page}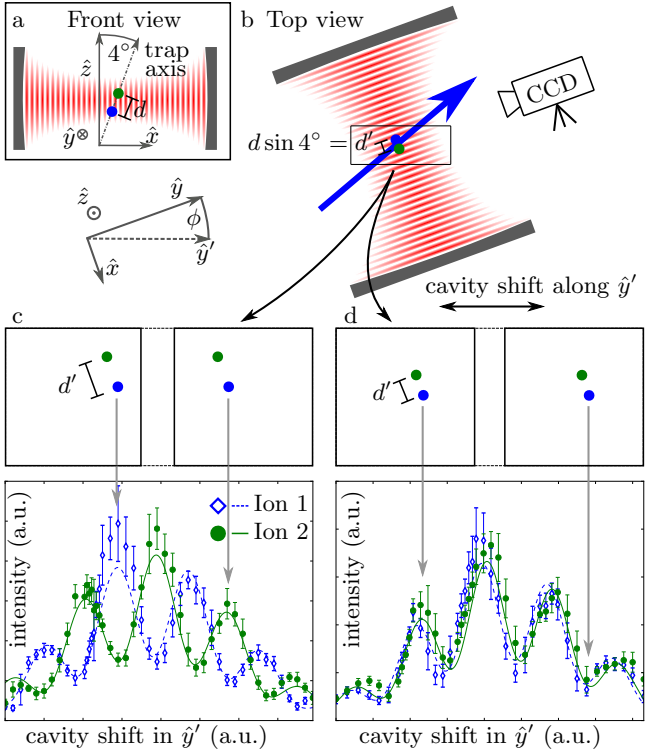
\caption{\label{fig_coupling} Two ions are trapped in a linear Paul trap within an optical cavity. The cavity axis is defined as $\hat{x}$. 
(a) There is a $4^\circ$ angle between the trap axis and $\hat{z}$. The separation $d$ between the ions can be tuned.
(b) A piezo stage translates the cavity along $\hat{y}'$. As $\hat{y}'$ is tilted at an angle $\phi \sim 5^\circ$ with respect to $\hat{y}$, the coupling of each ion to the TEM$_{00}$ mode is sinusoidally modulated.  The projection of $d$ in the $xy$ plane is $d'$.
(c)  For a projected ion--ion separation of $d' = 670$~nm, as the cavity is translated along $\hat{y}'$, the ions couple to the cavity with phase difference $0.9\pi$.
The cavity standing wave at 866~nm (red) is used to repump the ions, and fluorescence of the ions at 397~nm is measured on an EM-CCD.
(d)  For $d' = 370$~nm, the relative phase difference is $0.2\pi$.}
\end{figure}

Piezo stages allow us to shift the cavity with respect to a string of ions along both the cavity axis $\hat{x}$ and the near-orthogonal axis $\hat{y}'$ as indicated in Fig.~\ref{fig_coupling}.
Along the trap axis, the ions' relative and absolute positions can be shifted by adjusting voltages applied to the trap endcaps, which determine the axial confinement potential.
There is a $4^{\circ}$ angle between the trap axis and the $yz$ plane of the cavity mode (Fig.~\ref{fig_coupling}a).  
Due to a small angle $\phi$ between $\hat{y}$ and $\hat{y}'$ (Fig. \ref{fig_coupling}b),
the ions interact with a Gaussian TEM$_{00}$ mode modulated by the cavity standing wave as the cavity is translated along $\hat{y}'$ \cite{Stute12a}.

To determine the ions' positions with respect to the cavity mode, we rely on the fact that fluorescence on the $4^2S_{1/2} \leftrightarrow 4^2P_{1/2}$ $^{40}\text{Ca}^+$ transition at 397~nm requires a repump laser.
The ion is driven by a 397~nm laser from the side of the cavity and repumped by a cavity standing wave resonant with the $3^2D_{3/2} \leftrightarrow 4^2P_{1/2}$ transition, at $\lambda = 866$~nm.
The reflectivity of the cavity mirrors is optimized for this wavelength and for the nearby $3^2D_{5/2} \leftrightarrow 4^2P_{3/2}$ transition at 854~nm.
The standing-wave intensity is below saturation, so that the fluorescence of each ion depends on the ion's position in the standing wave.
A CCD camera images both ions.

An axial trap frequency of $2\pi \times 450$~kHz corresponds to a spacing of $d=9.6~\mu$m between two $^{40}\text{Ca}^+$ ions.  
The projection of this spacing along the cavity axis is given by $d' = d\sin{4^{\circ}} = 670~\text{nm} \approx 3\lambda/4$, sufficient to position one ion in a field node and the second ion in an antinode. 
This case is shown in  Fig. \ref{fig_coupling}c, in which the intensity of the cavity field seen by each ion is plotted as the cavity is shifted along $\hat{y}'$.
A separate calibration is used to translate the measured fluorescence at 397~nm  into intensity at 866~nm.
By fitting a sinusoidally modulated Gaussian to the data, we extract a relative phase of $0.9 \pi$ between the two ions with respect to the standing wave.


To couple two ions maximally to the cavity mode, we increase the axial trap frequency to $2\pi \times 1.09~$MHz, corresponding to $d =  5.3~\mu$m and $d' \approx \lambda/2$.  
This separation together with an appropriate cavity position allows us to position both ions in neighboring antinodes. 
In the situation shown in Fig. 1d, the two ions experience almost the same field as the cavity is translated; a phase difference of $0.2 \pi$ is determined from the fit.
More generally, this technique can be used to select any target phase difference between these two extremes.


\begin{figure}
\def\svgwidth{\columnwidth}
\executeiffilenewer{figure2.svg}{figure2.pdf}{inkscape -z -D --file=figure2.svg --export-pdf=figure2.pdf --export-latex --export-area-page}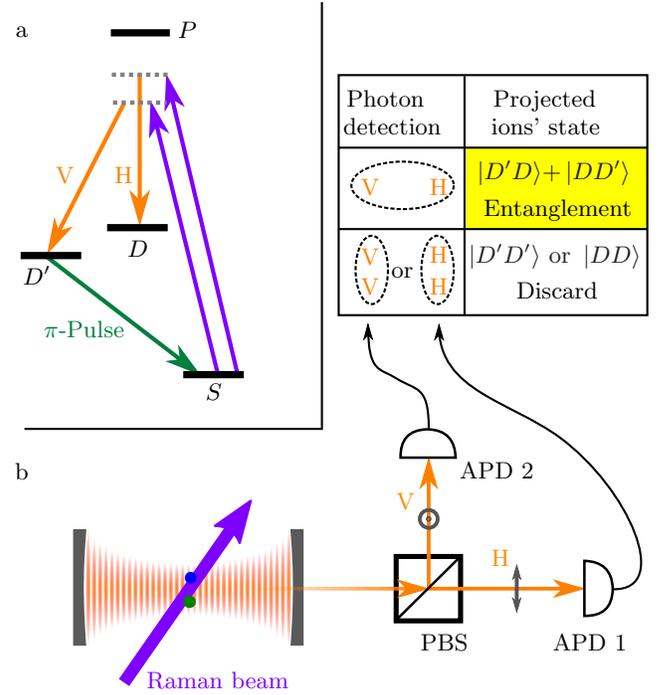
\caption{\label{fig2}
(a) Both ions are prepared in the electronic state $\ket{S}$. 
A Raman process, driven by a bichromatic field at 393~nm detuned from the excited state $\ket{P}$, generates two cavity photons at 854~nm. 
The photons' polarization, either horizontal ($H$) or vertical $(V)$, is entangled with the states $\ket{D}$ and $\ket{D'}$ of both ions. An optical $\pi$-pulse at 729~nm then coherently transfers population in $\ket{D'}$ to $\ket{S}$. 
(b) A polarizing beam splitter (PBS) at the cavity output enables measurement of polarization components using avalanche photodiodes (APDs).
Measurement of one $H$- and one $V$-polarized photon projects the ions into an entangled state.
 }
\end{figure}

We now use the second configuration, in which two ions are equally coupled to the cavity, to generate ion--ion entanglement.
Our entangling scheme relies on the method for ion-photon entanglement described in Ref. \cite{Stute12}.
Following optical pumping to $\ket{S} = \ket{4^2S_{1/2}, m_J=-1/2}$, a bichromatic Raman field at 393~nm drives the cavity-mediated transitions $\ket{S}\to\ket{D}  = \ket{3^2D_{5/2}, m_J=-3/2}$ and $\ket{S}\to\ket{D'}  = \ket{3^2D_{5/2}, m_J=-5/2}$, as shown in Fig. \ref{fig2}a.
Applying this Raman process to a single ion results in entanglement of the ion's electronic state with the polarization of a single cavity photon  at 854~nm:
\[ \ket{\psi} = \sqrt{1/2} ( |D H\rangle + |D'  V\rangle), \]
where a horizontally ($H$) or vertically ($V$) polarized photon is generated with equal probability.

Applying the Raman process to two ions coupled to the cavity generates two photons.
The two photons exit the cavity in the same spatial mode, providing a natural path interference.
$H$ and $V$ components  are split at a polarizing beamsplitter and detected at single-photon counters.
If one $H$ and one $V$ photon are detected (Fig. \ref{fig2}b), the state $ \ket{\Psi_{\text{tot}}} = \ket{\psi} \otimes \ket{\psi}$ is projected onto the state
\begin{equation*}
  \label{eq:entangleda}
 \ket{\Psi_{\text{herald}}} = \sqrt{1/2} ( \ket{DD'} + \ket{D'D}  ).
\end{equation*}
The joint detection event thus heralds ion--ion entanglement \cite{Moehring07}.
In order to perform state readout via fluorescence detection \cite{Leibfried03,Haeffner08},
we map $\ket{\Psi_{\text{herald}}}$ onto the qubit basis $\{\ket{S}, \ket{D}\}$ with a $\pi$ pulse on the $\ket{D'} \leftrightarrow  \ket{S}$ optical transition at 729~nm,
ideally generating the Bell state
\begin{equation*}
  \label{eq:entangled}
\ket{\Psi^+} = \sqrt{1/2} ( \ket{DS} + \ket{SD}  ).
\end{equation*}

The fidelity of the experimentally generated state \ket{\Psi} with respect to the target state \ket{\Psi^+}
can be bounded without reconstructing the full two-ion density matrix $\rho = \ket{\Psi}\bra{\Psi}$ \cite{Sackett00, Leibfried03a}.
Specifically, the fidelity is determined from three components of $\rho$:
\begin{align}
F_{\Psi^+} &= \bra{\Psi^+}  \rho  \ket{\Psi^+} \notag  \\
&=  (\rho_{SD,SD} +  \rho_{DS,DS})/2 + \mathrm{Re}(\rho_{SD,DS}).
\end{align}
The first term represents a direct measurement of population in states $\ket{SD}$ and $\ket{DS}$.
This population, equivalent to the probability that one ion is in $\ket{S}$, is determined by fluorescence detection on a photomultiplier over multiple trials.
More generally, the photomultiplier measurement allows us to determine $p_k$, the probability that $k$ ions are in $\ket{S}$, where $p_0 + p_1 + p_2 = 1$.

The second term of  $F_{\Psi^+}$ represents coherences between $\ket{SD}$ and $\ket{DS}$.
To estimate these coherences, we first implement two global $\pi/2$ rotations on the $\ket{D} \leftrightarrow  \ket{S}$ optical transition \cite{Benhelm08, Slodicka13}.
The first rotation $\sigma_x^{(1)} \sigma_x^{(2)}$ maps \ket{\Psi^+} to $\ket{\Phi^+} = \sqrt{1/2} (\ket{SS} + \ket{DD})$, where $\sigma_j^{(i)}$ denotes a Pauli spin operator acting on ion $i$ and $j=x,y,z$.
The second rotation is given by $\sigma_\phi^{(1)} \sigma_\phi^{(2)}$, where $\sigma^i_\phi =\sigma_x^i \cos\phi + \sigma_y^i \sin \phi$ and $\phi$ is the relative phase between the pulses.
The rotations are followed by a measurement of the parity $P$, defined as $p_0+ p_2 -  p_1$.
The parity oscillates as a function of $\phi$ and reaches a maximum for $\phi = \pi/2$ \cite{Slodicka13}, where
\begin{align}
P(\pi/2) = 2~\mathrm{Re}(\rho_{SD,DS} -\rho_{SS,DD})
\end{align}

Since $\rho_{SS,DD}$ may be nonzero, a second parity measurement is required, in which the first $\sigma_x^{(1)} \sigma_x^{(2)}$ rotation is not implemented. 
By measuring whether parity oscillations occur with contrast $C$, we can bound $\mathrm{Re}(\rho_{SS,DD})$ from above \cite{SuppMat}, thus bounding $F_{\Psi^+}$ from below. 
This bound is given by
\begin{align*}
F_{\Psi^+} \geq (\rho_{SD,SD} +  \rho_{DS,DS} + P(\pi/2) - C)/2.
\end{align*}

\begin{figure}
\includegraphics[width=0.5\textwidth]{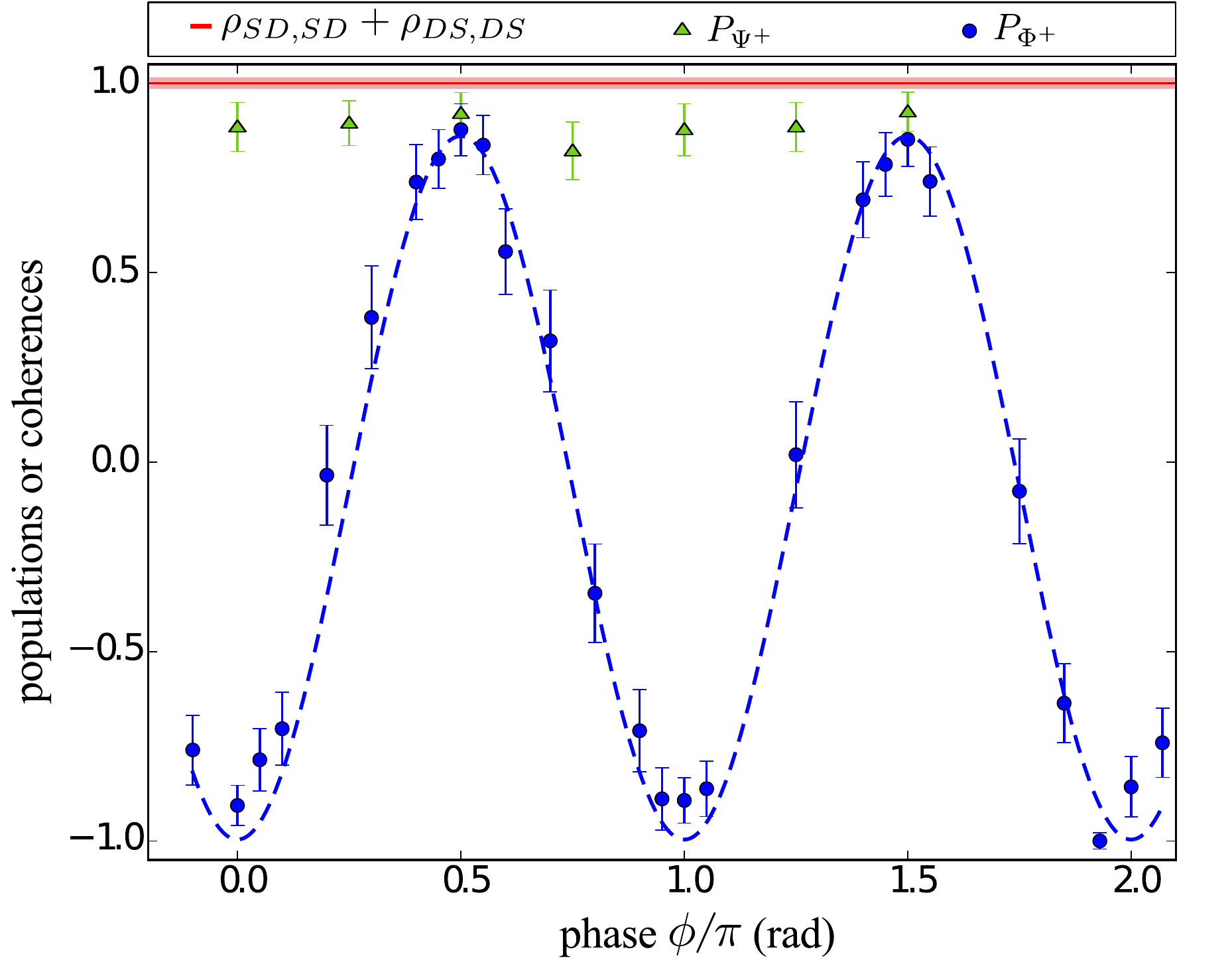}
\caption{\label{fig_parity}  Three measurements are used to bound the fidelity of the entangled state. First, the sum of population terms $\rho_{SD,SD}$ and $\rho_{DS,DS}$ is determined directly after entanglement. 
This measurement is independent of phase and is indicated by a red line whose width represents the uncertainty.
Second, after two $\pi/2$ rotations on the $\ket{S} \leftrightarrow \ket{D}$ optical transition, the parity is measured. 
The parity oscillates as a function of the relative phase of the rotations (blue circles). A sinusoidal fit is indicated by a dashed line.
Third, the parity is measured after only one $\pi/2$ pulse as a function of that pulse's phase (green triangles). 
Each data point represents about 50 entanglement events.  
Error bars represent one standard deviation, where the sources of error are projection noise and the determination of $p_k$ from fluorescence data \cite{SuppMat}.}
\end{figure}

In a single experimental sequence, the ions are first Doppler-cooled, then optically pumped to $\ket{S}$.
This preparation step lasts 1.7 ms.
Next, a 40~$\mu$s bichromatic Raman pulse is applied.
If two orthogonal photons are not detected at the APDs within these 40~$\mu$s, optical pumping to $\ket{S}$ and the Raman pulse are repeated up to ten times.
If all ten trials are ineffective, the sequence starts again with Doppler cooling.
If a first photon is detected at time $t_1$ and a second photon at time  $t_2$ after the Raman pulse is switched on, the mapping $\ket{D'} \rightarrow \ket{S}$ is implemented, and fluorescence detection for 2~ms determines how many ions are in $\ket{S}$.
For coherence measurements, analysis rotations are implemented before fluorescence detection.
For 25 values of the phase $\phi$, approximately 1000 entanglement events are recorded, corresponding to 1.5 hours of acquisition.

The data corresponding to a time interval $T=t_2 - t_1  \leq 0.5~\mu$s between photon detection events are plotted in Fig. 3.
The population measurement is indicated by a line, where $\rho_{SD,SD} +  \rho_{DS,DS}=1.00\pm0.03$.
After the $\sigma_x^{(1)} \sigma_x^{(2)}$ and $\sigma_\phi^{(1)} \sigma_\phi^{(2)}$ rotations, the parity $P_{\Phi^+}(\phi)$ oscillates with period $\pi$ and has a value of $0.86\pm0.01$ at phase $\pi/2$, determined from a sinusoidal fit.
A similar fit to the parity $P_{\Psi^+}(\phi)$ measured without the $\sigma_x^{(1)} \sigma_x^{(2)}$ rotation yields a contrast $C = 0.02 \pm 0.03$.
These three values result in a lower bound for the fidelity $F_{\Psi^+} \ge (91.9 \pm 2.5)\%$.

\begin{figure}
\includegraphics[width=0.5\textwidth]{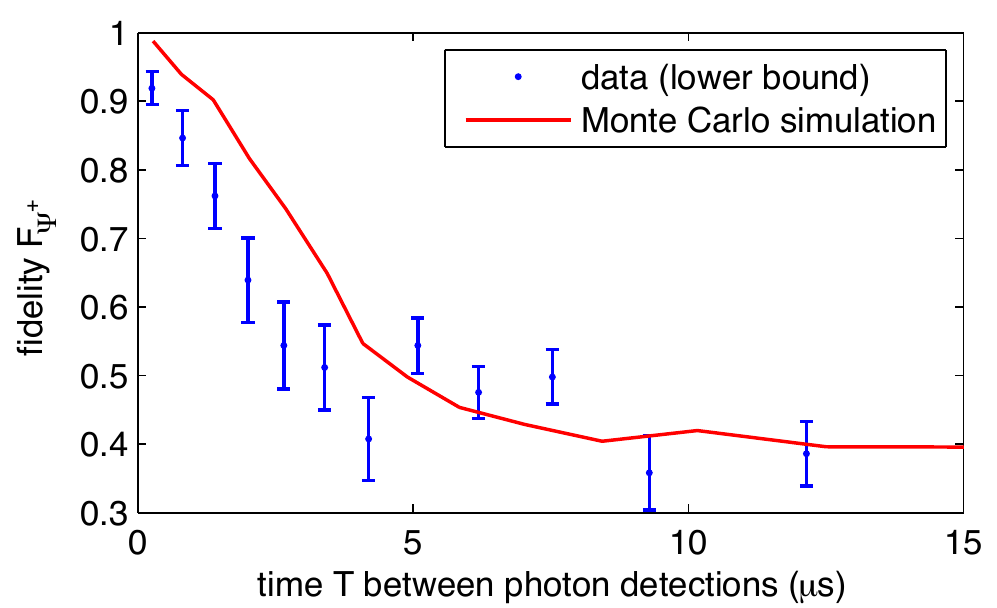}
\caption{\label{fig_decay} 
Lower bound of the fidelity of the entangled state with respect to $\Psi^+$ as a function of the detection interval $T$ between photons. 
As scattering within this interval removes the indistinguishability of the ions, the fidelity decreases with increasing $T$. 
The fidelity drops below 50\% for large $T$ due to the possibility of coherent evolution after a scattering event, which may result in the Bell state $\Psi^-$.
This process is reproduced by a Monte Carlo simulation of the fidelity (red line).
Error bars correspond to one standard deviation.
}
\end{figure}

The contrast of parity oscillations decreases with increasing $T$.
In Fig. \ref{fig_decay},  the lower bound for the fidelity $F_{\Psi^+}$ is plotted as a function of $T$ for the full data set.
Each time bin contains $\sim$1750 entanglement events, so that the first bin corresponds to the data of Fig. \ref{fig_parity}; the bin spacing increases with $T$ as photon coincidence becomes less likely.
The observed loss of coherence is due to scattering from the $4^2P_{3/2}$ manifold back to $\ket{S}$ while the coherent Raman transition is in progress.
A scattering event allows us in principle to distinguish between the two ions.
Thus, the indistinguishability required for entanglement is lost.

One might expect that for large times $T$, the coherence $\rho_{SD,DS}$ approaches zero with population $p_1$ remaining constant, resulting in a fidelity at the classical limit of 50\%.
However, the fidelity drops below this limit (Fig. \ref{fig_decay}), due to the generation of coherences following scattering.
When an ion spontaneously decays after one cavity photon has already been detected, both ions are projected to a state with one ion in $\ket{S}$ and the other in $\ket{D}$ or $\ket{D'}$.
The second photon can be generated in two ways.
In one process, the second photon is generated from the ion in $\ket{S}$ and exits the cavity.
In another process, the photon does not exit the cavity but is reabsorbed via the reverse Raman transition by the ion in $\ket{D}$ or $\ket{D'}$.
Subsequently, this ion emits another photon, which exits the cavity.
Because a geometric phase is acquired in the second process, the interference of these two processes results in a negative  $\rho_{DS,SD}$.

This effect has been reproduced in numerical simulations via the quantum Monte Carlo method
\cite{Tan99}.
For each ion, the system Hamiltonian takes into account the four electronic levels shown in Fig. 2a and an additional Zeeman state in the $D$ manifold that is weakly coupled to $\ket{S}$ by an off-resonant Raman process.   
The ions are coupled to the two cavity modes.
Collapse operators correspond to the decay terms $\kappa$ and $\gamma$ and to a drive laser linewidth of 10~kHz.
The two-ion state is evaluated as a function of the arrival times of two orthogonally polarized photons, with the result shown in Fig.~\ref{fig_decay}.

In comparison to the data, the simulated fidelities are systematically higher for $T < 5$~ms, indicating that coherence between states $\ket{SD}$ and $\ket{DS}$ is lost more quickly than expected.  
We attribute these faster decoherence rates in the experiment to laser frequency noise that may be underestimated in the simulations.
Other sources of experimental imperfections which may decrease the fidelity in the percent regime include imperfect state readout and state preparation due to the finite ion temperature, detector dark counts, and atomic decoherence from magnetic field fluctuations.

In the experiment presented here, we have generated ion--ion entanglement with a fidelity of at least $(55 \pm 2)\%$ at a rate of 4.3 events per second, while the high-fidelity data subset of Fig. \ref{fig_parity} 
corresponds to a rate of  0.2 events per second.
These rates would be similar if the ions were located in spatially separated cavities and can be improved by implementing faster cooling \cite{Lin13} and a cavity with higher output efficiency and faster cavity decay. 
The present scheme could also be extended to generate $N-$ion Dicke states, heralded by the detection of $m$ horizontal and $(N-m)$ vertical photons \cite{Duan03}.
We further note that control of the coupling of multiple ions to the cavity mode constitutes an important step toward hybrid quantum networks, in which small ion-trap registers in cavities are linked via optical fibers \cite{Jiang07}.
Remote ions could be coupled to one another by shifting each register within its cavity, with additional ions available for error correction or storage, resulting in a scalable resource for quantum computation.

\begin{acknowledgments}
We thank L. Slodi\v{c}ka and C. Roos for helpful discussions.
We gratefully acknowledge support from the Austrian Science Fund (FWF):  Project. Nos. F4003 and F4019, 
the European Research Council through the CRYTERION Project,
the European Commission via the Atomic QUantum TEchnologies (AQUTE) Integrating Project, 
and the Institut f\"ur Quanteninformation GmbH.
\end{acknowledgments}

\appendix*
\section{Supplemental Material}

\subsection{Cavity parameters}

The cavity-field decay rate is $\kappa = 2\pi \times 50$ kHz, and the sum of atomic decay channels from the $ 4^2P_{3/2}$ state is $\gamma = 2\pi \times 11.5$~MHz.
The atom-cavity coupling strength on the $3^2D_{5/2} \leftrightarrow 4^2P_{3/2}$ $^{40}\text{Ca}^{+}$ transition is $g_0 =  2 \pi \times 1.4$~MHz.
Due to the Gaussian mode cross-section, for two ions separated by $5.3~\mu$m along the trap axis, the maximum coupling strength for each ion is $2 \pi \times 1.3$~MHz.
As in Ref.~\cite{Stute13}, from comparison of the data with simulations, we infer a reduced coupling strength $g_{\text{motion}} =  2 \pi \times 1.0$~MHz due to thermal motion and drifts of the ions' position from the cavity antinode.
This value is slightly higher than in Ref. \cite{Stute13}, which we attribute to improved alignment of the ions with respect to the cavity mode.

The effective coupling strengths $g_{i,\text{eff}}$ and atomic decay rates $\gamma_{i,\text{eff}}$ for each of the two drive fields can be calculated from Eqs. 2 and 3 of Ref. \cite{Stute12a}, where $i$ is an index that labels the drive field.  Note that  $g_{i,\text{eff}}$ is half of the effective Rabi frequency  $\Omega_{\text{eff}}^i$ defined in Ref. \cite{Stute12a}.  The drive-field Rabi frequencies are $\Omega_1 = 47$~MHz and $\Omega_2 = 29$ MHz, where $\Omega_1$ couples to $D$ and $\Omega_2$ to $D'$.  Both fields are detuned by $\approx 400$~MHz from  the $ 4^2P_{3/2}$ state.  The values $g_{\text{eff}}$ and $\gamma_{\text{eff}}$ given in the text for a bichromatic field are calculated as
\begin{align*}
g_{\text{eff}} &= \sqrt{g_{1,\text{eff}}^2 + g_{2,\text{eff}}^2}  \\
\gamma_{\text{eff}}&= \gamma_{1,\text{eff}} + \gamma_{2,\text{eff}}
\end{align*} 
since $g_{i,\text{eff}} \propto \Omega_i$,  $\gamma_{i,\text{eff}} \propto \Omega_i^2$, and the sum of two Rabi frequencies is the square root of the sum of their squares.

\subsection{Fidelity calculation}

Eq. 1 in the main text expresses the fidelity $F_{\Psi^+}$ of the experimentally generated state with respect to the maximally entangled Bell state $\ket{\Psi^+}$ as a function of population and coherence terms.
However, the coherence term $\rho_{SD,DS}$ in Eq. 1 can not be extracted directly from the measurement represented by Eq. 2, as the term $\rho_{SS,DD}$ also appears here.

$\rho_{SS,DD}$ represents the coherence between $|SS\rangle$ and $|DD\rangle$.  The coherence terms in a density matrix corresponding to a physical state must fulfill the condition
\[
|  \rho_{SS,DD} |  = | \langle SS | \rho  | DD\rangle  |\le \sqrt{ \rho_{SS,SS} } \sqrt{ \rho_{DD,DD} }.
\]


As described in the main text, a fluorescence measurement allows the independent determination of probabilities $p_0, p_1$, and $p_2.$
The probabilities $p_0$ and $p_2$ correspond to $\rho_{DD,DD}$ and $\rho_{SS,SS}$, respectively.
For the particular case in which $p_0=p_2=0$, the coherence term $\rho_{SS,DD}=0$, and therefore $\rho_{SD,DS}$ can be directly determined.

In general, $\rho_{SS,DD}$ may be nonzero.  In order to bound its value, an additional measurement of the parity operator is required.
After joint detection of orthogonal photons and the subsequent mapping from $|D'\rangle \to |S\rangle$, the rotation $\sigma_\phi^{(1)} \sigma_\phi^{(2)}$ is implemented.
The phase $\phi$ is defined with respect to the phase of the mapping pulse.
The parity $P(\phi)$ is measured as function of $\phi$.
A function $P_{\text{fit}}(\phi) = C \sin (2 \phi + \phi_0)$ can be fit to the data of the resulting parity oscillation, where $C=2 | \mathrm{Re} (\rho_{SS,DD})| $ and $\phi_0$ is the phase that maximizes the contrast in order to obtain a maximum bound \cite{Benhelm08}.

Since
\[
\mathrm{Re} ( \rho_{SD,DS} - \rho_{SS,DD} )    \le  \mathrm{Re}(\rho_{SD,DS}) +  |\mathrm{Re}(\rho_{SS,DD})|,
\]
the second term of Eq. 1 
can now be bounded from above:
\begin{equation*}
\label{bound}
\mathrm{Re} (\rho_{SD,DS})  \ge   \mathrm{Re} ( \rho_{SD,DS} - \rho_{SS,DD})  - C/2,
\end{equation*}
where $\mathrm{Re} ( \rho_{SD,DS} - \rho_{SS,DD} )$ is determined from the measurement of Eq. 2.  Thus, the fidelity $F_{\Psi^+}$ can be bounded from below.

\subsection{Fluorescence statistics}

After ion--ion entanglement is heralded and the ion states are mapped and rotated, fluorescence detection is implemented for 2~ms.
Each ion in the state $|S\rangle$ scatters photons, which are recorded by a PMT.
Ideally, the histogram of detected counts is described by a sum of Poissonian distributions with mean values $\mu_n=n \cdot d$, where $d$ is the mean number of photons that a single ion scatters in 2~ms and $n$ is the number of ions in $|S\rangle$.

In the measurements presented, efficient Doppler cooling of the ions was achieved with two laser beams, since one beam is not sufficient to address both radial motional modes.
These two beams are also used for fluorescence detection.
Interference between these two fields at the ions' position leads to fluctuations of the fluorescence, such that $d$ is not longer constant.
Thus, the statistics of the detected counts do not take the form of a Poissonian distribution.
Instead, for $n$ ions in $|S\rangle$,  the statistics are well described by a sum of Gaussian distributions
\[
g_{n}(s)= \frac{1}{\sigma_n \sqrt{2\pi}} \cdot e^{-(s-s_n)^2/2\sigma_n^2},
\]
in which a single fluorescence detection results in $s$ counts, and $\sigma_n$ and $s_n$ represent the variance and the mean value of the distribution.
This follows from the central limit theorem, given the assumption that the fluctuations of $\mu_n$ are Gaussian.

In order to find $s_n$ and $\sigma_n$, all data are analyzed, including those in which no coincidences occur or photons of identical polarization are detected.

Given $s$ counts, the goal of each fluorescence detection is to identify $n$, the number of ions in $\ket{S}$, as well as the error in this identification.
We find the value of $i \in \{ 0,1,2 \}$ that maximizes $g_i(s)$ and assign $n = i$.
Following Ref. \cite{Myerson08}, the associated error $\delta$ is calculated as
\[
\delta = \frac{ g_{n}(s)}{\sum_i g_{i}(s)}.
\]

A set of measurements consists of $\eta_0$ events in which we assign $n = 0$, $\eta_1$ events corresponding to $n=1$, and $\eta_2$ events corresponding to $n=2$.
The probabilities $p_0, p_1$, and $p_2$ are then calculated as
\[
p_i = \frac{\eta_i}{\sum_i \eta_i}.
\]
The uncertainty $\delta p_i$ associated with $p_i$ has two components:
\[
\delta p_i = \delta p_{\text{stat}} + \delta p_{\text{proj}}.
\]
The first component,  $\delta p_{\text{stat}}$, is the result of propagating the errors $\delta$ from all $\eta_i$ events.
The second component corresponds to the quantum projection noise \cite{PhysRevA.47.3554}
\[
\delta p_{\text{proj}} =  \sqrt { \frac { \big ( 1-p_i \big ) p_i }{\eta_i}  }
\]
for $p_i \ne 0,1 $ and
\[
\delta p_{\text{proj}} =\sqrt {\eta_i }
\]
otherwise.

\bibliography{cqed_bibsonomy,SM}

\end{document}